\documentclass[reprint,twocolumn,superscriptaddress,amsmath,amssymb,amsfonts,aps,prb,floatfix]{revtex4-2}
\usepackage{times}
\usepackage{booktabs}
\usepackage{graphicx}
\usepackage{dcolumn}
\usepackage{float}
\usepackage[makeroom]{cancel}
\usepackage{bm}
\usepackage{ulem}
\usepackage[colorlinks=true, citecolor=blue, linkcolor=blue, urlcolor=blue]{hyperref}
\usepackage{bbold}
\DeclareMathAlphabet\mathbfcal{OMS}{cmsy}{b}{n}
\usepackage{wasysym}
\usepackage{amssymb}
\usepackage{amsmath}
\usepackage{siunitx}

\usepackage[usenames,dvipsnames]{xcolor}

\begin{document}
% \title{Nematicity, Quantum Geometry, and Superconductivity in Crystalline Graphene}
\title{Nematic Enhancement of Superconductivity in Multilayer Graphene via Quantum Geometry}

\author{Gal Shavit}
\affiliation{Department of Physics and Institute for Quantum Information and Matter, California Institute of Technology,
Pasadena, California 91125, USA}
\affiliation{Walter Burke Institute of Theoretical Physics, California Institute of Technology, Pasadena, California 91125, USA}
% \author{Jason Alicea}
% \affiliation{Department of Physics and Institute for Quantum Information and Matter, California Institute of Technology,
% Pasadena, California 91125, USA}

\begin{abstract}

Multilayer graphene materials have recently emerged as a fascinating versatile platform for correlated electron phenomena, hosting superconductivity, fractional quantum Hall states, and correlated insulating phases. 
A particularly striking experimental observation is the recurring correlation between nematicity in the normal state -- manifested by spontaneous breaking of the underlying $C_3$ symmetry -- and the stabilization of robust superconducting phases.
Despite its ubiquity across different materials, devices and experiments, this trend has thus far lacked a clear microscopic explanation. 
In this work, we identify a concrete mechanism linking nematic order to enhanced superconductivity. 
We demonstrate that $C_3$-symmetry breaking strongly reshapes the Bloch wavefunctions near the Fermi level, producing a pronounced enhancement and redistribution of the so-called quantum metric.
This effect drastically amplifies superconducting pairing mediated by the quantum geometric Kohn–Luttinger mechanism
[G. Shavit \textit{et al.}, \href{https://journals.aps.org/prl/abstract/10.1103/PhysRevLett.134.176001}{Phys. Rev. Lett. 134, 176001 (2025)}].
% ~\cite{Shavit_Alicea_PhysRevLett.134.176001}.
Our analysis reveals that nematicity naturally boosts the superconducting coupling constant in experimentally relevant density regimes, providing a compelling explanation for observed correlations.
These results establish the central role of geometric effects in graphene superconductivity and highlight nematicity as a promising avenue for engineering stronger unconventional superconducting states.

\end{abstract}
\maketitle

\section{Introduction}

Multilayer graphene materials have been at the forefront of correlated quantum matter in recent years.
The immense theoretical interest they garner stems from remarkable experimental progress and discoveries, including
rich fractional quantum Hall landscapes~\cite{Bernalfqh1_Zibrov2017,bernalfqh2_doi:10.1126/science.aao2521,bernalfci3_doi:10.1126/science.aan8458,bernalfqh4_junzhu_PhysRevX.12.031019,YuvalRonen_BLG_FQH_Kumar2025},
abundance of correlated phases in moir\'e devices~\cite{TBG1_CaoCorrelatedInsulator,TBG_2CaoUnconventionalSC,KimTrilayerGrapheneSuperconductivity,Park2021StronglyCoupledSuperconductivityTrilayer,Park2022MATngFamily,matngYIRAN,TDBG1_Shen2020,TDBG_folk_Su2023,Quasicrystal_Uri2023},
emergence of robust quantum anomalous Hall states
\cite{hBNgoldhaberGordon,hBNyoung,BilayerMonolayerChern2,LongJu_experiment_lu2024extendedquantumanomaloushall,PentaG2_Han2024,Young_FQAH_choi2024electricfieldcontrolsuperconductivity},
and a myriad of fractional Chern insulators
~\cite{PENTA_FCILu2024,fci_hexa_chinese}.

One of the most intriguing phenomena is the appearance of superconductivity at low enough temperatures in crystalline graphene materials~\cite{RTGsuperconductivityZhou2021,ZhouYoungBLGZeeman,nadj_ISOC_BBGBLGZhang2023,Nadj_zhang2024twistprogrammablesuperconductivityspinorbitcoupled,Young_patterson2024superconductivityspincantingspinorbit,BBG_electronside_li2024tunable,young_nadj_BBG_RTG_SC,LongJuyang2024diverseimpactsspinorbitcoupling,LONGJU_4_5_chiral_sc_Han2025,JIALI_Hexa_nguyen2025hierarchytopologicalsuperconductingstates,Yankowitz_octo_hepta_kumar2025superconductivitydualsurfacecarriersrhombohedral,xiaomeng_hexa_deng2025superconductivityferroelectricorbitalmagnetism}.
The ubiquity of superconductivity in these devices fueled vigorous theoretical activity, proposing possible pairing mechanisms, and attempting to account for various idiosyncrasies in experimental observations~\cite{DasSarmaBLGphonons,RTG_Ashvin,annular_PhysRevLett.127.247001,AlejandroPacoBLG,StonerBlockadeBBGPhysRevB.108.024510,dong2025superconductivityspincantingfluctuationsrhombohedral,Paco_RTG_PhysRevB.105.075432,PhysRevB.107.104502,KWANBLG,LiangFugeier2024chiraltopologicalsuperconductivityisospin}.
One experimental trend, however, has thus far received little 
% to no 
theoretical attention: The correlation between nematicity in the normal state, i.e., spontaneous breaking of the microscopic $C_3$ symmetry, and the appearance of more robust superconducting states.
The trend is summarized in Table~\ref{table:thetable}, establishing the aforementioned correlation over many devices and several experimental groups.

\begin{figure}
    \centering
    \includegraphics[width=8cm]{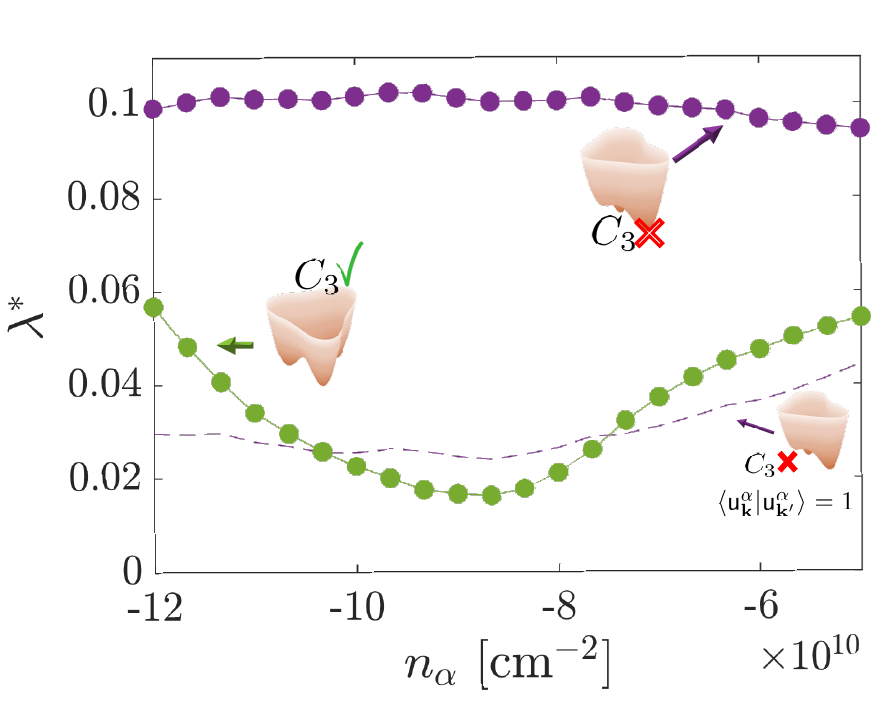}
    \caption{
    \textbf{Nematicity enhances quantum geometric superconductivity in bilayer graphene.}
    Plot of the maximal superconducting coupling constant $\lambda^*$ (see Sec.~\ref{sec:QGpairing}) as a function of single-flavor density $n_\alpha$.
    Green dots represent the case for $C_3$-symmetric normal state (three equal energy trigonal warping pockets in the band structure), and purple represents the nematic normal state (two out of three pockets below the Fermi level).
    Dashed purple is for the nematic state, in the absence of quantum geometric underscreening, i.e., geometric form-factors in Eq.~\eqref{eq:staticpol} are all set to unity.
    Notice that the superconducting critical temperature $T_c\propto e^{-1/\lambda^*}$, such that the differences in $T_c$ can be enormous.
    The parameters used in this figure (see text for details):
    $U=60$ meV,
    $d=20$ nm,
    $\epsilon=4$.
    \label{fig:mainlambda}}
\end{figure}

\begin{table}
\begin{tabular}{SSS} \toprule
    {Description} & {$T_c^{\rm normal}$ [mK]} & { $T_c^{\xcancel {C_3}}$ [mK]}  \\ \midrule
    {BLG, Device A \cite{ZhouYoungBLGZeeman}}  & {n/a} & 26  \\
    {BLG, Device B \cite{ZhouYoungBLGZeeman}}  & {n/a} & 22  \\
    \midrule
    {BLG + WSe$_2$, Device 1 \cite{nadj_ISOC_BBGBLGZhang2023}}  & {n/a} & 260  \\
    {BLG + WSe$_2$, Low ISOC \cite{Nadj_zhang2024twistprogrammablesuperconductivityspinorbitcoupled}}  & {n/a} & 150  \\
    {BLG + WSe$_2$, Mid ISOC \cite{Nadj_zhang2024twistprogrammablesuperconductivityspinorbitcoupled}}  & {40} & 450  \\
    {BLG + WSe$_2$, High ISOC \cite{Nadj_zhang2024twistprogrammablesuperconductivityspinorbitcoupled}}  & {60} & 500  \\
     {BLG + WSe$_2$, electron side \cite{BBG_electronside_li2024tunable}}  & {n/a} & 210  \\
     {BLG + WSe$_2$, hole side \cite{BBG_electronside_li2024tunable}}  & {n/a} & 400  \\
     {BLG + WSe$_2$, \cite{young_nadj_BBG_RTG_SC}}  & {30} & 255  \\
     \midrule
    {R3G + WSe$_2$  \cite{Young_patterson2024superconductivityspincantingspinorbit}}  & {n/a} & 285  \\
    {R3G + WSe$_2$, electron side  \cite{LongJuyang2024diverseimpactsspinorbitcoupling}}  & {n/a} & 186  \\
    {R3G + WSe$_2$, hole side  \cite{LongJuyang2024diverseimpactsspinorbitcoupling}}  & {n/a} & 104  \\ \bottomrule
\end{tabular}
\caption{
\textbf{Superconductivity in crystalline graphene is correlated with nematicity.}
A survey of experimental superconductivity discoveries in Bernal bilayer graphene devices (BLG) and rhombohedral trilayer graphene (R3G).
Most of the reported superconductors have a Tungsten diselenide substrate (WSe$_2$) which stabilizes superconductivity by inducing Ising spin-orbit coupling (ISOC) in the graphene layers.
Here, $T_c^{\rm normal}$ is the maximal superconducting critical temperature for regions in the phase diagram where the normal state is likely $C_3$-symmetric, 
whereas $T_c^{\xcancel {C_3}}$ is the corresponding critical temperatures for regions which appear incompatible with this symmetry.
We omitted from this table superconductors whose ``Fermiology'', extracted from magnetic quantum oscillations, were inconclusive as to the breaking of the $C_3$ symmetry (e.g., the R3G superconducting domes in Ref.~\cite{RTGsuperconductivityZhou2021}).
} \label{table:thetable}
\end{table}

The origin of spontaneous $C_3$ symmetry breaking is best understood in Bernal bilayer graphene (BLG) and rhombohedral trilayer graphene (R3G) devices~\cite{donglevitov_PhysRevB.107.075108,MacdonaldRTG}.
When an electric displacement field is applied in a direction perpendicular to the graphene planes, the band structure at low densities consists of three small trigonal warping pockets.
Coulomb interactions may then favor the occupation of one or two of these pockets, instead of the kinetically-favored equally distributed occupation.
In fact, even if the three small pockets are not fully formed, which is likely the case in stacks with higher number of layers, a Pomeranchuk-type instability can still be triggered by the interactions, leading to a nematic state~\cite{macdonald_crescent_PhysRevB.91.155423}.
Experimentally, the appearance of nematicity was inferred by examination of the Fermi-surface-induced quantum oscillations in high quality devices.
In the relevant density regimes, if the number of degenerate small Fermi surface is not a multiple of three (referring to the existence of spin and valley flavors) -- $C_3$ can be reasonably assumed to be broken.
Recently, non-linear transport measurements on crystalline graphene devices purported a more direct observation of this effect~\cite{JIA_LI_nematicity_BLG_lin2023spontaneousmomentumpolarizationdiodicity,JIALI_r6g_morissette2025coulombdrivenmomentumspacecondensation}.

In this work, we set forth a concrete mechanism which naturally gives rise to strong correlations between nematicity and superconductivity in these systems.
The mechanism relies on the fact that the Bloch wavefunctions of the active band get strongly modified by the $C_3$-symmetry breaking order parameter.
In other words, one needs to go beyond heuristic models of momentum-space polarization (e.g., three-pocket models~\cite{donglevitov_PhysRevB.107.075108}), and consider the implications of concrete microscopic order parameters.

We find that in the experimentally relevant density regimes there is a pervasive and marked increase of the so-called quantum metric in the vicinity of the Fermi energy when $C_3$ is broken.
More to the point, the nematic order parameter renders the wavefunctions of the occupied Bloch states vastly different from the unoccupied states.
In turn, this leads to a giant enhancement of superconductivity mediated by the quantum geometric Kohn-Luttinger pairing effect (see Fig.~\ref{fig:mainlambda}), recently discussed in Refs.~\cite{Shavit_Alicea_PhysRevLett.134.176001,Shizeng_jahin2025enhancedkohnluttingertopologicalsuperconductivity}.
By solving this experimental puzzle, we provide a compelling argument as to the relevance of the unconventional geometric pairing mechanism to the phenomenon of multilayer graphene superconductivity.

The rest of the paper is organized as follows.
In Sec.~\ref{sec:QGpairing} we survey the essential details of the superconducting mechanism mentioned above, and how it relates to the quantum geometric properties of the band hosting the paired electrons.
Sec.~\ref{sec:nematicitygeometry} is devoted to illustrating the connection between the nematic order parameter and significant changes to the quantum geometric properties of the band in graphene multilayers.
We bring these two components together in Sec.~\ref{sec:BLGworkedout}, and directly demonstrate how nematicity strongly enhances superconductivity in BLG (where the empirical correlations are the most obvious).
We conclude our findings in Sec.~\ref{sec:discussionconclusions}, where we highlight their implications for future exploration of superconductivity in correlated quantum materials.

\section{Quantum geometric pairing}\label{sec:QGpairing}
We begin by briefly recapitulating the basic principles underlying quantum geometric Kohn-Luttinger superconductivity~\cite{Shavit_Alicea_PhysRevLett.134.176001,Shizeng_jahin2025enhancedkohnluttingertopologicalsuperconductivity}.
The important mechanism at play is the so-called \textit{geometric underscreening}: 
The electronic system poorly screens interactions with high momentum transfer due to non-trivial band quantum geometry.
This band property renders particle-hole fluctuations inefficient at mediating the screening, once their Bloch wavefunctions are sufficiently dissimilar from one another. 

Let us substantiate this mechanism by considering a system of interacting electrons, projecting its Hamiltonian onto a set of $N_f$ active bands, labeled by the flavor index $\alpha$.
The projection operator onto one of these bands at momentum $\bf k$ is $P_{\bf k}^\alpha=|{\mathsf u}_{\bf k}^\alpha\rangle\langle {\mathsf u}_{\bf k}^\alpha|$, where $|{\mathsf u}_{\bf k}^\alpha\rangle$ are the Bloch wavefunctions of the $\alpha$ band.
The projected interacting Hamiltonian is
\begin{equation}
    H = \sum_{{\bf k},\alpha} \left(\epsilon_{{\bf k},\alpha}-\mu\right)\psi^\dagger_{{\bf k},\alpha}\psi_{{\bf k},\alpha}
    +
    \frac{1}{2A}\sum_{\bf q} V_{\bf q}\Tilde{\rho}_{\bf q}\Tilde{\rho}_{\bf -q}, \label{eq:projectedBareHamiltonian}
\end{equation}
where $\psi_{{\bf k},\alpha}$ is a fermionic annihilation operator of an electron in flavor $\alpha$ and momentum $\bf k$, $\mu$ is the chemical potential, $\epsilon_{{\bf k},\alpha}$ are the band dispersions, $V_{\bf q}$ is the repulsive density-density interaction, and $A$ is the area of the system.
The projected density operator is
\begin{equation}
    \tilde{\rho}_{\bf q}=\sum_{\alpha,{\bf k}} \Lambda^\alpha_{\bf k,k+q}\psi^\dagger_{{\bf k+q},\alpha}\psi_{{\bf k},\alpha}\,,
    \,\,\,\,\,
    \Lambda^\alpha_{\bf k,k'}= \left\langle {\mathsf u}_{\bf k}^\alpha|{\mathsf u}_{\bf k'}^\alpha\right\rangle.
    \label{eq:projecteddensity}
\end{equation}
% Hereafter we assume that $H$ preserves time-reversal symmetry; the $N_f$ flavors then come in pairs, such that each flavor $\alpha$ admits a time-reversed partner $\bar{\alpha}$, which together satisfy
% $\epsilon_{{\bf k},\alpha}=\epsilon_{-{\bf k},\bar{\alpha}}\equiv \epsilon_{\bf k}$ and
% $\Lambda^\alpha_{\bf k,k'}=\left(\Lambda^{\bar{\alpha}}_{\bf k,k'}\right)^*\equiv\Lambda_{\bf k,k'}$.
Henceforth, we will consider for concreteness two-dimensional systems with dual-gate screened Coulomb interactions,
$V_{\bf q}=\frac{2\pi e^2}{\epsilon}\frac{\tanh{qd}}{q}$, where $q=\left|{\bf q}\right|$, $\epsilon$ is the dielectric constant and $d$ is the distance to the gates.

The bare repulsion $V_{\bf q}$ is further screened in the random phase approximation (RPA) by the aforementioned particle-hole fluctuations of the normal-state Fermi liquid,
\begin{equation}
    V_{\bf q}^{\rm RPA}=\frac{V_{\bf q}}{1+\Pi_{\bf q}  V_{\bf q}},\label{eq:Vrpa}
\end{equation}
\begin{equation}
    \Pi_{\bf q}= - \sum_{\alpha,{\bf k}}
    \left|\Lambda^\alpha_{\bf k,k+q}\right|^2
    \frac{ n\left(\xi^\alpha_{\bf k+q}\right) - n\left(\xi^\alpha_{\bf k}\right) }
    {\xi^\alpha_{\bf k+q} - \xi^\alpha_{\bf k}}.
    \label{eq:staticpol}
\end{equation}
Here, $\Pi_{\bf q}$ is the static polarization bubble, $\xi^\alpha_{\bf k}=\epsilon^\alpha_{\bf k}-\mu$ is the distance from the Fermi level and $n\left(x\right)=\left(1+e^{x/T}\right)^{-1}$ is the Fermi-Dirac distribution at temperature $T$.

The quantum geometric underscreening effect is captured by the dependence of the screening strength on the form-factors $\Lambda^\alpha_{\bf k,k+q}$ in Eq.~\eqref{eq:staticpol}.
One simple case which showcases the quantum geometric effects is the scenario where all bands are degenerate and have parabolic dispersion, $\epsilon^\alpha_{\bf k}=\frac{\left|{\bf k}\right|^2}{2m}$.
The small momentum dependence of the screening in this case (taking the $T\to0$ limit for concreteness),
\begin{equation}
    \Pi_{\bf q} \approx \Pi_0
    \left[
    1-q_\mu q_\nu \left\langle{g}_{\mu\nu}\right\rangle_{\rm FS}
    \right] + O\left(q^3\right),\label{eq:overscreenedPi}
\end{equation}
where $\Pi_0=N_f\frac{m}{2\pi}$, and
\begin{equation}
    g_{\mu \nu} = {\rm Tr}\left[\partial_\mu P_{\bf k} \partial_\nu P_{\bf k}\right] \label{eq:FSmetric}
\end{equation}
is the Fubini-Study metric.
Fermi-surface averaging is denoted by $\left\langle\cdot\right\rangle_{\rm FS}$.

Notably, the geometric effect at small momenta is negative, i.e., screening is suppressed at higher momentum.
Considering Eq.~\eqref{eq:Vrpa}, the effective repulsion may acquire complicated momentum dependence, and in particular may \textit{increase with momentum} in certain regimes.
This sort of interaction would in turn promote superconductivity with an order parameter that changes sign along the Fermi surface.

Using the screened interaction as the mediator of superconducting pairing, which takes place between bands $\alpha$ and $\bar{\alpha}$ (with degenerate dispersion
$\xi^\alpha_{\bf k}=\xi^{\bar{\alpha}}_{\bf k}\equiv\xi_{\bf k}$), 
the self-consistent Bardeen-Cooper-Schreiffer (BCS) gap equation is
\begin{equation}
    \Delta_{\bf k}=
    -\int\frac{d{\bf k'}}{\left(2\pi\right)^2}
    u_{\bf kk'}
    \frac{\tanh{\frac{E_{\bf k'}}{2T}}}{2E_{\bf k'}}
    \Delta_{\bf k'}.\label{eq:fullBCS}
\end{equation}
Here, the Bogoliubov spectrum is $E_{\bf k}=\sqrt{\xi_{\bf k}^2+\left|\Delta_{\bf k}\right|^2}$, and the interaction picks up a secondary quantum-geometric contribution,
\begin{equation}
    u_{\bf kk'}=\Lambda^\alpha_{\bf k,k'}\Lambda^{\bar\alpha}_{\bf -k,-k'}
    V_{\bf k-k'}^{\rm RPA}.\label{eq:ukkp}
\end{equation}
We note that if $\alpha$ and $\bar\alpha$ are time-reversal partner bands, the form factors in~\eqref{eq:ukkp} can be replaced by $\left|\Lambda^\alpha_{\bf k,k'}\right|^2$.

In order to capture the salient features of quantum-geometry-enabled superconductivity in this work, it would prove convenient to approximate the full BCS gap equation by its linearized form in the vicinity of the Fermi surface,
\begin{equation}
    \Delta_p
    =-\log\frac{W}{T_c}\int_{\rm FS}
    \frac{dp'}{\left(2\pi\right)^2}
    \frac{1}{v_{p'}}
    u_{p-p'}
    \Delta_{p'},\label{eq:linearBCS}
\end{equation}
where $T_c$ is the critical temperature, $W\sim E_F$ is introduced as an energy cutoff, $\int_{\rm FS}$ represents integration over the Fermi surface, and $v_{p'}$ is the Fermi velocity at point $p'$ on the Fermi surface.
One obtains Eq.~\eqref{eq:linearBCS} from~\eqref{eq:fullBCS}
by replacing $d{\bf k'}\to dp_\perp dp'$, linearizing the spectrum near the Fermi surface $\xi_{\bf k}\to v_{p'}p_\perp$, and integrating over the perpendicular coordinate $p_\perp$.
Discretizing in momentum space, Eq.~\eqref{eq:linearBCS} may be thought of as a linear equation system, characterized by the matrix
${\cal M}_{p,p'}=-
    \frac{\sqrt{\delta p \delta p'}}{\left(2\pi\right)^2}
    \frac{1}{\sqrt{v_p v_{p'}}}
    u_{p-p'}$, 
where $\delta_p$ is the momentum resolution at point $p$.
Eigenfunctions $\phi^i_p$ of the matrix $\cal M$ with positive eigenvalues $\lambda^i$ correspond to different superconducting channels with critical temperatures $T_c^i\approx W\exp\left(-1/\lambda^i\right)$.
The highest positive eigenvalue $\lambda^*$ then marks the leading superconducting instability.

\section{Nematicity and quantum geometry in graphene multilayers}\label{sec:nematicitygeometry}

\subsection{Bernal bilayer graphene}

Let us first establish the connection between the spontaneous development of nematicity in Bernal bilayer graphene and significant changes of the active band quantum geometry.
The low-energy continuum model expanded around one of the valley points of graphene (the Hamiltonian in the other valley can be obtained by time-reversal) may be written as~\cite{mccannElectronicPropertiesBilayer2013}
\begin{equation}
    {\cal H}_{\rm BLG}=
    \sum_{\bf k} \Psi^\dagger\left({\bf k}\right)h_{\rm BLG}\left({\bf k}\right)\Psi\left({\bf k}\right),\label{eq:hblgschematic}
\end{equation}
\begin{equation}
    h_{\rm BLG}\left({\bf k}\right) = 
    \begin{pmatrix}
    U/2 &  v_0 \Pi^\dagger & -v_4 \Pi^\dagger & -v_3 \Pi \\
    v_0 \Pi & \Delta' +U/2 & \gamma_1 & -v_4 \Pi^\dagger \\
    -v_4 \Pi & \gamma_1 & \Delta' -U/2 & v_0 \Pi^\dagger \\
    -v_3 \Pi^\dagger & -v_4 \Pi & v_0 \Pi & -U/2
    \end{pmatrix},\label{eq:hblgspecific}
\end{equation}
where $\Pi = k_x + i k_y$ and $v_i \equiv \frac{\sqrt{3} a}{2} \gamma_i$ ($a=0.246$ nm).
Eq.~\eqref{eq:hblgspecific} is written in the basis corresponding to electron annihilation operators of the form 
$\Psi=\left(\psi_{1A},\psi_{1B},\psi_{2A},\psi_{2B}\right)^T$ ($\psi_{is}$ is a fermionic annihilation operator in layer $i$ and sublattice $s$). 
The term $U$ accounts for a displacement-field-induced interlayer potential difference.
In this work, we adopt the parameters from Ref.~\cite{jungAccurateTightbindingModels2014}, $\gamma_0 = 2.61$~eV, $\gamma_1 = 361$ meV, $\gamma_3 = 283$ meV, $\gamma_4 = 138$ meV, and $\Delta' = 15$ meV.

\begin{figure}
    \centering
    \includegraphics[width=8.5cm]{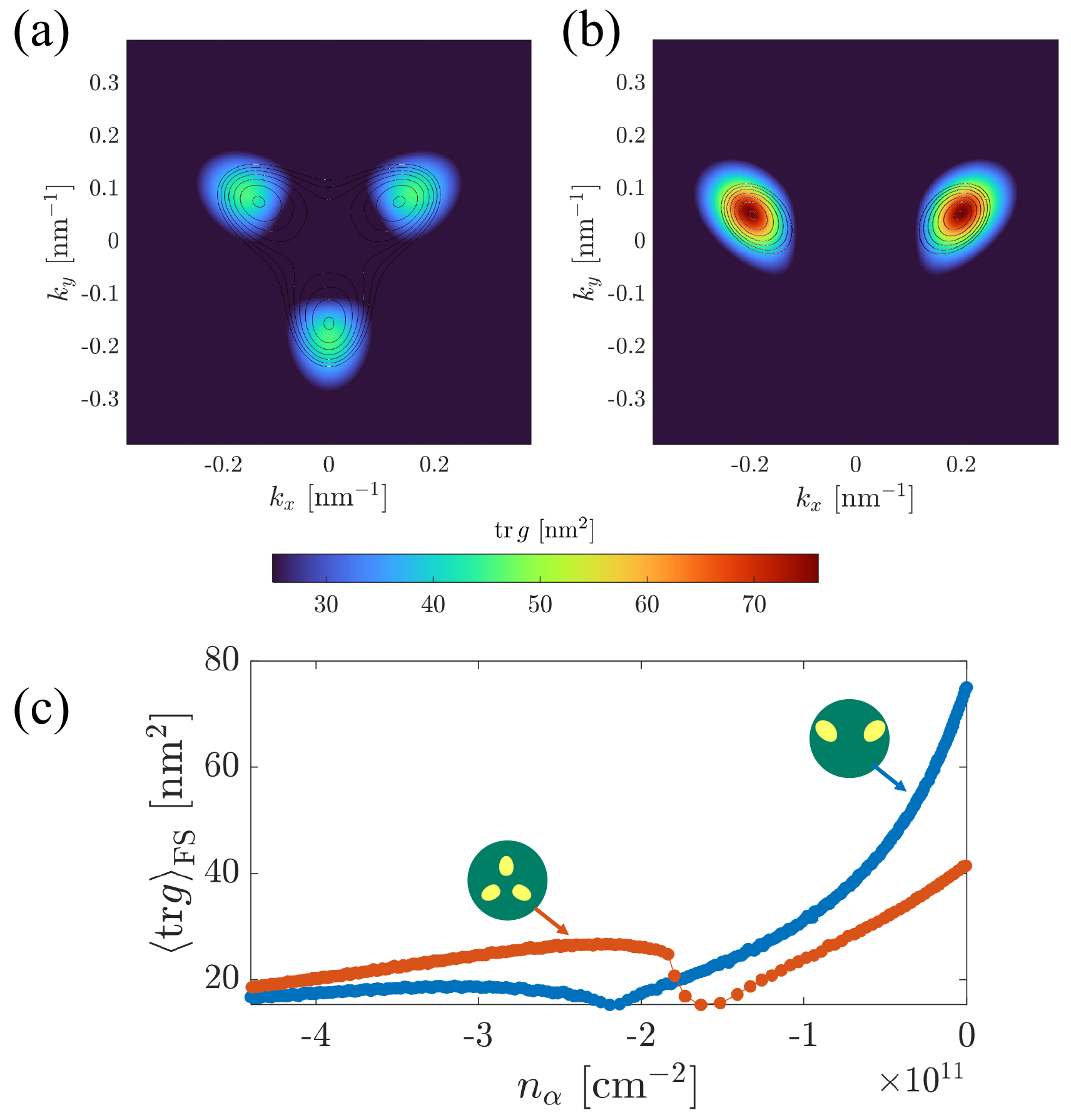}
    \caption{
    \textbf{Quantum metric and $C_3$ breaking in BLG.}
    (a)
    Trace of the quantum metric plotted for the lowest lying valence band of Bernal bilayer graphene, Eqs.~\eqref{eq:hblgschematic}--\eqref{eq:hblgspecific}.
    (b)
    The same quantity plotted for the scenario with $C_3$ symmetry breaking [see Eq.~\eqref{eq:blgnemmodification}], where a total of two out of three pockets are occupied. (Here, $\phi=\phi_{2\rm p}=\pi$, and $\epsilon_{\rm nem}=20$ meV.)
    For both panels the black contours represent the Fermi surfaces at the top of the band, plotted in 1 meV intervals, and we use $U=60$ meV.
    (c)
    Fermi-surface averaged quantum metric as a function of single-flavor density $n_\alpha$ for the normal case appearing in panel (a) (red), and for the nematic state in panel (b) (blue).
    \label{fig:BLGgeometry}}
\end{figure}

We focus henceforth on the lowest-lying valence band of the BLG Hamiltonian, where both superconductivity and $C_3$ breaking phases were most prevalently recorded.
In the presence of large $U$, the top of this valence band is comprised of three hole pockets connected by a 120-degree rotation around the valley point.
Notably, the quantum metric associated with this band shares the same $C_3$ symmetry, yet its concentration is somewhat offset from the center of these pockets, as demonstrated in Fig.~\ref{fig:BLGgeometry}a.

Nematicity may be introduced as a single-particle order parameter to the matrix in Eq.~\eqref{eq:hblgspecific}.
For concreteness, let us consider the modification
\begin{equation}
    h_{\rm BLG}\left({\bf k}\right)\to h_{\rm BLG}\left({\bf k}\right)+\begin{pmatrix} &  &  & \epsilon_{{\rm nem}}e^{i\phi}\\
\\
\\
\epsilon_{{\rm nem}}e^{-i\phi}
\end{pmatrix},\label{eq:blgnemmodification}
\end{equation}
which explicitly breaks the $C_3$ symmetry of the inter-sublattice interlayer hopping term of the Hamiltonian (proportional to $v_3$).
We stress that the modification above, introduced in a single-particle form, originates in spontaneous symmetry breaking due to Coulomb interactions.
Although Eq.~\eqref{eq:blgnemmodification} could arise from strain intrinsic to the device, this scenario is not consistent with the experimental observations of abrupt changes of the number of Fermi surfaces in the normal state (e.g., from three to two per flavor) as a function of density or electric displacement field.

The phase $\phi$ associated with this additional term determines the direction at which the resulting band dispersion is ``tilted'' in momentum space.
In practice, since the spontaneous symmetry breaking facilitates disparate population of the three trigonal warping pockets, $\phi$ is expected to be an integer multiple of $\pi/3$.
Without loss of generality, one may  consider the restricted cases $\phi_{1\rm p}=0$, and $\phi_{\rm 2p}=\pi$.
Under the convention of $\epsilon_{\rm nem}>0$, these favor the occupation of a total of one or two pockets, respectively.

As illustrated in Fig.~\ref{fig:BLGgeometry}b, this sort of order parameter has a profound impact on the distribution of quantum metric in the active band.
The effect is two fold: (i) The quantum metric is greatly enhanced, and (ii) the centers of the Fermi pockets now lie much closer to where the metric is strongly concentrated.
Furthermore, in Fig.~\ref{fig:BLGgeometry}c we plot the Fermi surface average of the quantum metric as the relevant valence band is gradually filled with holes.
It is evident that at low densities the occupied hole states in the nematic phase are much more ``geometrically charged'' -- the quantum metric substantially increases in the occupied regions.
As we will demonstrate in the next Section, due to the geometric underscreening mechanism elaborated on in Sec.~\ref{sec:QGpairing}, this change of quantum geometric properties may lead to a giant enhancement of the superconducting tendencies in this band.

\subsection{Rhombohedral $n$-layer graphene}
The intimate relationship between $C_3$ symmetry breaking and enhanced quantum geometry in the vicinity of the Fermi level extends to the entire family of rhombohedral-stacked multilayer graphene.
This may be demonstrated in a simplified model, retaining only the intralayer Dirac dispersion of the electrons, alongside the purely-vertical interlayer hopping.
Projecting into the two lowest lying bands in the basis of the spinor 
$\Psi_{\rm RnG}=\left(\psi_{1A},\psi_{nB}\right)^T$
% ($\psi_{is}$ corresponds to a fermionic annihilation operator in layer $i$ and sublattice $s$
, 
the projected Hamiltonian in the vicinity of the $K$-valley point is~\cite{bernevig2025berrytrashcanmodelinteracting},
\begin{equation}
    {\cal H}_{{\rm RnG}}=\sum_{{\bf k}}\Psi_{{\rm RnG}}^{\dagger}\left({\bf k}\right)\begin{pmatrix}D_{{\bf k}} & t_{{\bf k}}\left(\pi_{{\bf k}}^{*}\right)^{n}\\
t_{{\bf k}}\pi_{{\bf k}}^{n} & -D_{{\bf k}}
\end{pmatrix}\Psi_{{\rm RnG}}\left({\bf k}\right),\label{eq:rngHamiltonian}
\end{equation}
where 
$\pi_{{\bf k}}=\frac{v\left(k_{x}+ik_{y}\right)}{\gamma}$,
$t_{{\bf k}}=-\gamma\sqrt{\frac{1-\left|\pi_{{\bf k}}\right|^{2}}{1-\left|\pi_{{\bf k}}\right|^{2n}}}$,
$D_{{\bf k}}=U\left[-\frac{n-1}{2}+\frac{\left(n-1\right)\left|\pi_{{\bf k}}\right|^{2n+2}+\left|\pi_{{\bf k}}\right|^{2}-n\left|\pi_{{\bf k}}\right|^{2n}}{\left(1-\left|\pi_{{\bf k}}\right|^{2}\right)\left(1-\left|\pi_{{\bf k}}\right|^{2n}\right)}\right]$,
$v\approx542$ meV$\cdot$nm is the Dirac cone Fermi velocity,
$\gamma\approx355$ meV is the interlayer hopping amplitude,
and $U$ is the interlayer potential difference.

\begin{figure}
    \centering
    \includegraphics[width=9cm]{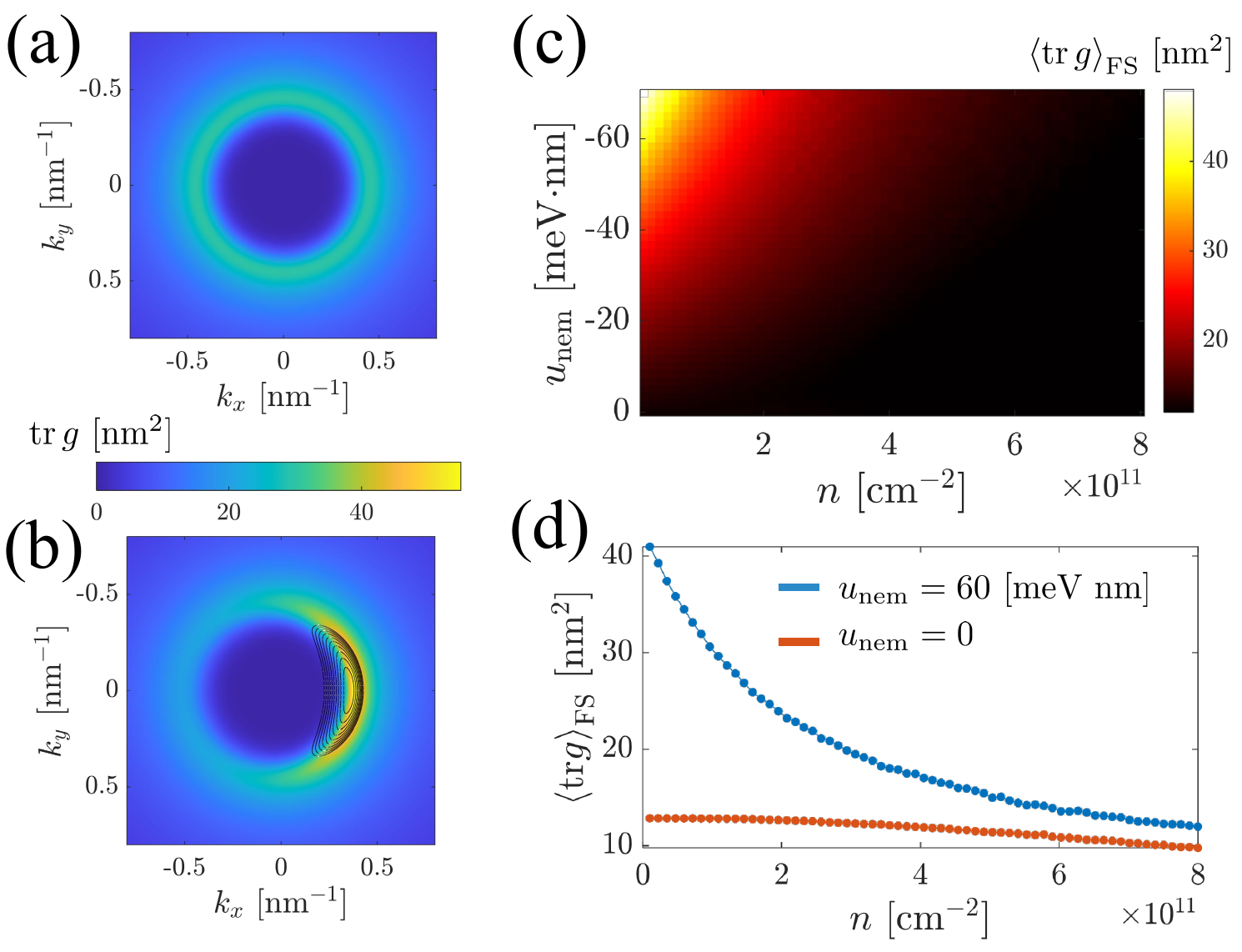}
    \caption{
    \textbf{Quantum geometry in symmetry-broken RnG.}
    (a)
    Trace of the quantum metric plotted for the conduction band with $n=5$ layers.
    The quantum metric is peaked in the shape of a rotationally symmetric ring.
    (b)
    The quantum metric redistributes when nematicity takes place.
    It gets both enhanced and concentrated in a narrow sector of the Brillouin zone.
    For illustrative purposes, the black contours depict the Fermi surface at the bottom of the band, where the Fermi energy is modified by consecutive 1 meV intervals.
    Here, $u_{\rm nem}=60$ meV nm.
    (c)
    Fermi-surface averaged quantum metric as a function of electronic density and symmetry-breaking order parameter.
    The nematicity-geometry correlation is apparent.
    (d)
    Two horizontal cuts of panel (c), with values corresponding to panels (a) and (b).
    Throughout this figure  we use $U=25$ meV.
    \label{fig:RNGgeometry}}
\end{figure}

This Hamiltonian results in two relatively flat bands at small momentum, separated by a large gap.
Since we neglect any trigonal warping terms, the bands of this Hamiltonian are actually fully rotationally invariant around the $K$-valley point in momentum space.
Notably, the quantum metric (and also the Berry curvature) is strongly peaked in the shape of a ring, see Fig.~\ref{fig:RNGgeometry}a.
The location of the ring can be inferred from the competition between the diagonal and off-diagonal terms in Eq.~\eqref{eq:rngHamiltonian}.
At small momentum, the diagonal interlayer potential difference dominates, and the wavefunction is mostly layer-polarized.
At large enough momenta, the off-diagonal term dominates, and the wavefunction has approximately equal weights on both extremal layers.
When the two terms in the Hamiltonian are comparable, there occurs a significant change in the nature of the electronic wavefunction, leading to an enhancement of quantum metric.

We may artificially break the rotational symmetry by modifying the Hamiltonian in Eq.~\eqref{eq:rngHamiltonian},
$D_{\bf k}\to D_{\bf k}+u_{\rm nem}{\bf k}\cdot\hat{\bf d}$, where $\hat{\bf d}$ is a unit vector explicitly breaking the rotational symmetry.
Besides ``tilting'' the electronic bands towards the direction dictated by $\hat{\bf d}$, this term redistributes the quantum geometry as well, as evident from Fig.~\ref{fig:RNGgeometry}b.
Notably, at low densities, the occupied electrons reside in a region of enhanced quantum metric.
This is demonstrated in Fig.~\ref{fig:RNGgeometry}c-d, where we plot the Fermi surface averaged quantum metric, $\left\langle{\rm tr}\,g\right\rangle_{\rm FS}
$, as a function of density, for different strength of the symmetry breaking term.
Notice that the nematic transition induces a similar trend in RnG as it did for the BLG model (see Fig.~\ref{fig:BLGgeometry}c).
Namely, the quantum metric becomes significantly stronger in the lowest-lying parts of the bands, i.e., closer to the $n=0$ charge neutrality point.

Recall that the screening of the Coulomb interaction, a crucial component of the superconducting mechanism discussed here, is mediated by particle-hole fluctuations.
These fluctuations are weighted by the wavefunction overlaps, as explained above.
If the occupied electronic states reside in a region of large quantum metric, the aforementioned overlaps are necessarily further diminished at higher momentum transfer.
This is the epitome of quantum geometric underscreening, which plays the key role in promoting Kohn-Luttinger superconductivity.

\section{Nematicity-driven superconductivity in BLG}\label{sec:BLGworkedout}
Having established how the spontaneous breaking of $C_3$ symmetry enhances the quantum geometric properties near the Fermi level for the relevant density regimes, let us now employ the framework detailed in Sec.~\ref{sec:QGpairing} and explore its effect on superconductivity.
Given that the bulk of empirical evidence points towards strong nematicity-superconductivity correlation in BLG devices, this is where we shall focus our efforts.

To bring out the full extent of this effect, and to better facilitate comparison between the nematic and $C_3$-symmetric scenarios, we concentrate on only one Fermi surface in each case.
In the normal case, this would be one out of three pockets (or possibly a multiple of three due to spin/valley flavor degeneracies), and in the nematic case, one pocket out of two (or multiple of two).
We set aside considerations such as inter-pocket interactions~\cite{KWANBLG} and interband pairing or scattering due to correlation/substrate-driven transitions~\cite{Nadj_zhang2024twistprogrammablesuperconductivityspinorbitcoupled}.
These considerations likely play a role in experimental scenarios, yet are expected to have a similar effect on the properties and strength superconductivity for the $C_3$-symmetric and nematic states.
Therefore, their inclusion (which requires additional assumptions in our modeling) would not alter the conclusions of this work, and are left for future theoretical studies of the experimental observations.

We parametrize the Fermi surface by an angle $\theta\in\left[0,2\pi\right]$, characterizing each point with respect to the center of the pocket.
We then discretize the Coulomb and RPA-screened interactions into interaction matrices,
$V_{\bf q}\to V_{\theta,\theta'}$,
$V^{\rm RPA}_{\bf q}\to V^{\rm RPA}_{\theta,\theta'}$,
such that the interaction between Fermi-surface segments at angles $\theta$ and $\theta'$ is determined by their momentum difference
${\bf q}= \left|{\bf k}_\theta-{\bf k}_{\theta '}\right|$.
This allows us to readily construct the appropriate matrices ${\cal M}_{\theta,\theta'}$ discussed in Sec.~\ref{sec:QGpairing}, and extract their leading eigenvalues.
These results are plotted in Fig.~\ref{fig:mainlambda}, showing substantial increase in the coupling constant in the nematic case.
The exponential dependence of the superconducting $T_c$ on these effective coupling constants thus signals colossal enhancement of superconductivity by the nematic normal state.

Crucially, the marked differences between the superconducting coupling constants may be attributed almost exclusively to quantum geometrical effects.
Repeating the steps of the calculations in the nematic case, with artificially setting all form factors in Eq.~\eqref{eq:overscreenedPi} to unity, reveals that the so called nematic advantage is entirely gone (dashed purple line in Fig.~\ref{fig:mainlambda}).

\begin{figure}
    \centering
    \includegraphics[width=9cm]{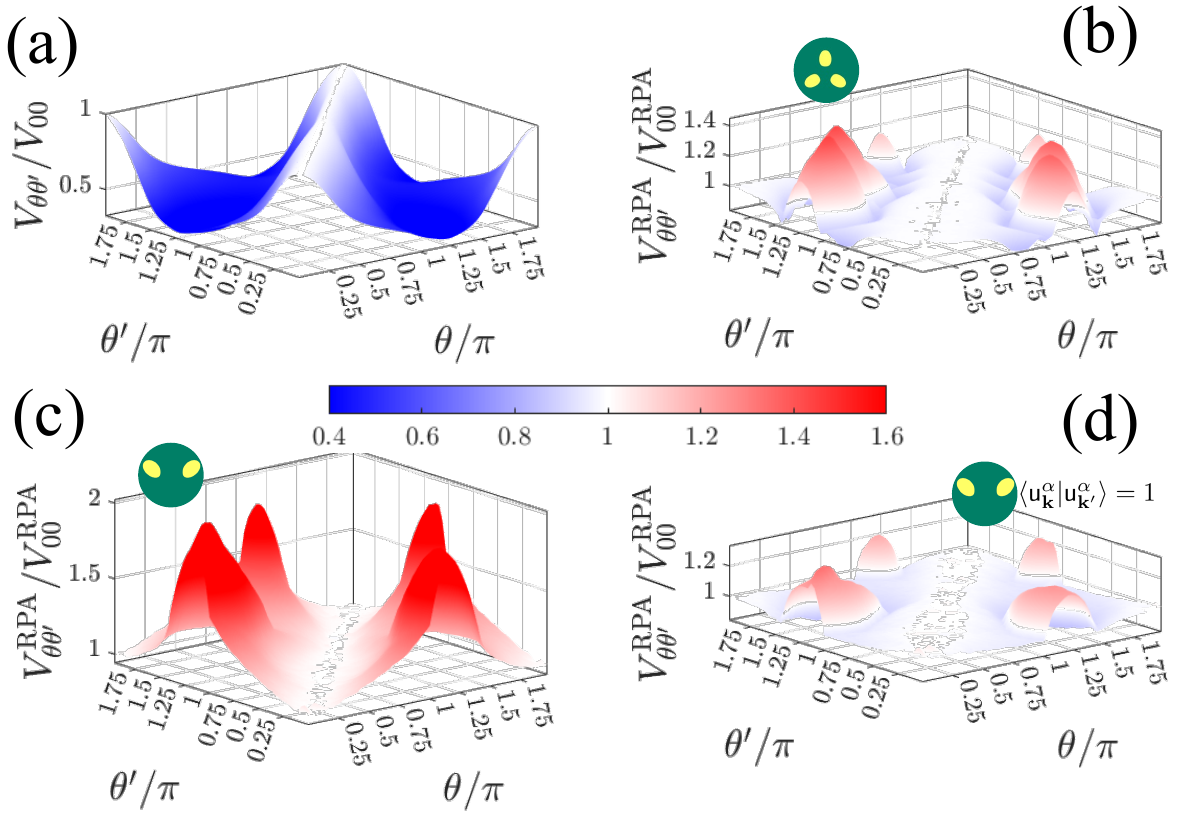}
    \caption{
    \textbf{Quantum geometric underscreening in BLG.}
    (a)
    Bare Coulomb repulsion between different segments of the Fermi surface at one of the trigonal warping pockets in BLG.
    (b)
    The RPA screened interaction in one of the Fermi pockets of the $C_3$-symmetric normal state.
    (c)
    The RPA screened interaction For the nematic state.
    (d)
    Same as (c), with all form-factors set to unity in the calculation of $\Pi_{\bf q}$ [Eq.~\eqref{eq:staticpol}].
    For all panels, the interactions are normalized such that the value on the diagonal of the interaction matrix is unity.
    Here, $n_\alpha=10^{11}$ cm$^{-2}$, $\epsilon=4$, and $d=20$ nm.
    For the nematic plots we used $\phi=\phi_{2\rm p}=\pi$ and $\epsilon_{\rm nem}=20$ meV.
    \label{fig:vrpa}}
\end{figure}

To better understand the cause for this effect, let us explore in more detail the (screened) interactions, represented in Fig.~\ref{fig:vrpa}.
In Fig.~\ref{fig:vrpa}a, we plot the \textit{bare} Coulomb repulsion between segments of one of the Fermi surfaces.
Notice an expected trend -- the diagonal entries are maximal, a consequence of the Coulomb repulsion decaying with increasing momentum separation  $\left|{\bf q}\right|$.
For the screened interactions, that is not necessarily the case.
In Fig.~\ref{fig:vrpa}b, the $V^{\rm RPA}_{\theta,\theta'}$ matrix of the $C_3$-symmetric case is plotted, showing ``flatter'' behavior near the diagonal, with some enhancement of the repulsion (as compared to the diagonal, not to the bare repulsion) between certain distant regions of the Fermi surface.
These regions, separated by large $\bf q$, experience an underscreened interaction, which in turn favors pairing instabilities in non s-wave channels (see further discussion in Ref.~\cite{Shavit_Alicea_PhysRevLett.134.176001}).

For the nematic normal state, illustrated in Fig.~\ref{fig:vrpa}c, the outcomes of this geometric underscreening are far more extreme.
Now the interaction on the diagonal, corresponding to small momentum transfer, is a pronounced \textit{minimum} of repulsion, and the repulsion is much more strongly peaked at large momentum difference.
Importantly, the marked differences between Fig.~\ref{fig:vrpa}b and Fig.~\ref{fig:vrpa}c are almost entirely due to quantum geometric effects.
By artificially negating these effects again, i.e., setting all the form factors to unity for the calculation of the screened interaction, we recover a screened interaction in Fig.~\ref{fig:vrpa}d that closely resembles the tame form of screening shown in Fig.~\ref{fig:vrpa}b.
This accounts for the trends seen in Fig.~\ref{fig:mainlambda}.

\section{Discussion}\label{sec:discussionconclusions}

In this work, we attempt to shed light on the observed strong correlation between a nematic ($C_3$ symmetry broken) normal state and robust superconductivity in graphene multilayers.
We propose that this correlation strongly points towards the quantum geometry of the active electronic bands as playing a major role in the formation and stabilization of superconducting domes in the phase diagram.

We demonstrated how the nematic transition naturally and generically leads to enhancement of the geometric quantum metric in the vicinity of the low-density-region Fermi surfaces.
By virtue of the quantum geometric underscreening mechanism (introduced in Sec.~\ref{sec:QGpairing} and explored more thoroughly in Ref.~\cite{Shavit_Alicea_PhysRevLett.134.176001}), a Kohn-Luttinger type of superconducting mechanism gets massively enhanced by this reconstruction of the Bloch wavefunctions.

This study concentrated on unconventional superconductivity arising from the nematic normal state in graphene multilayers.
The pairing scenario detailed above is distinct from previously explored nematicity-related mechanisms which rely on quantum-critical fluctuations near the spontaneous phase transitions~
\cite{Kivelsonia_nematic_PhysRevLett.114.097001,Islam2024_chubukov_fese_nematic}.
The latter fluctuation-mediated mechanism seems less likely given the experimental observations.
Specifically, the superconducting dome in the nematic phase may appear away from the nematic transition, see Refs.~\cite{young_nadj_BBG_RTG_SC,Young_patterson2024superconductivityspincantingspinorbit,LongJuyang2024diverseimpactsspinorbitcoupling}.
Occasionally, the superconducting dome is terminated sharply on the nematic side, highly consistent with the static nematicity-induced scenario presented above.

By bringing the strong empirical evidence for nematicity and its relation to superconductivity to the forefront, the case for a quantum-geometric mechanism for unconventional superconductivity has thus become significantly more compelling.
To the best of our knowledge, the scenario proposed in this work represents the first theoretical attempt at relating $C_3$-symmetry breaking and the superconducting critical temperatures in crystalline graphene multilayers.
The relative success of such an attempt hints at the origin of superconductivity in these materials, and further motivates a search for other correlated materials with highly non-trivial quantum geometric properties.
These hold promise for discoveries of novel superconducting phases.

We note that the geometric effects we explore here may be even more expansive than they appear at first glance.
The correlation between Ising-type spin orbit coupling (ISOC) and enhanced superconductivity has long been noted~\cite{nadj_ISOC_BBGBLGZhang2023,Young_patterson2024superconductivityspincantingspinorbit} (it is apparent even in Table~\ref{table:thetable}).
An extensive numerical Hartree-Fock study~\cite{ISOC_promotes_nema_PhysRevB.109.035113} has recently remarked that extrinsic ISOC (e.g., introduced by a WSe$_2$ substrate) enhances nematic tendencies in R3G.
Our work can thus provide a missing link between this observation and its effect on superconductivity, further strengthening the ISOC-superconductivity relation (currently understood exclusively through the lens of interband interactions in the Cooper channel~\cite{Nadj_zhang2024twistprogrammablesuperconductivityspinorbitcoupled,dong2025superconductivityspincantingfluctuationsrhombohedral}).

A direct consequence of the mechanism we explore here to relate nematicity and superconductivity is that superconductivity may be enhanced and become more pervasive throughout the phase diagram by controlled application of uniaxial strain.
This experimental knob has been recently implemented with some success in a moir\'e graphene device~\cite{ma2025giantelastoresistancemagicangletwisted}.
Application of strain in the optimal directions (corresponding to populating/depopulating the trigonal warping pockets) would definitively distinguish between the nematicity-driven-superconductivity scenario we espouse here, from any possible coincidental scenarios.

One may also consider an alternative form of explicit $C_3$ symmetry breaking, the application of an in-plane magnetic field.
To leading order, the effect of this single-particle perturbation is quite similar to the application of strain, yet in a way that breaks time-reversal symmetry: the momentum space band ``tilt'' acts differently on the two valleys~\cite{ratchet_blg_inplane_magnetic_PhysRevB.94.165404}.
One thus expects the magnetic field to have an adverse impact on superconductivity due to intervalley orbital depairing~\cite{StonerBlockadeBBGPhysRevB.108.024510,young_nadj_BBG_RTG_SC}.
Yet, if the depairing effect is somewhat compensated by the geometric-boost explored here, non-trivial dependence of the superconducting $T_c$ on the applied field strength may surface~\cite{ZhouYoungBLGZeeman}.
More curious still, if the Cooper pairs from within a single valley, as alleged in Ref.~\cite{LONGJU_4_5_chiral_sc_Han2025}, the orbital depairing could be suppressed, and a magnetic field (at the right in-plane angle) could enhance superconductivity by our prescribed quantum-geometric mechanism.
We note that a nematic state, and its intimate connection to in-plane magnetic fields, have been recently proposed as possible explanations for anomalous behavior observed in R9G~\cite{r9g_crescent_li2025transdimensionalanomaloushalleffect}, rendering our conclusions even more relevant to future explorations.

\begin{acknowledgments}
We thank Jason Alicea, Tobias Holder, \'Etienne Lantagne-Hurtubise, and Julian May-Mann for insightful discussions.
We especially thank Stevan Nadj-Perge and Yiran Zhang for bringing the issue of nematicity-superconductivity correlation in experiments to our attention.
We acknowledge support from the Walter Burke Institute for Theoretical Physics at Caltech, and from the Yad Hanadiv Foundation through the Rothschild fellowship. 
This work was performed in part at Aspen Center for Physics, which is supported by National Science Foundation grant PHY-2210452.

\end{acknowledgments}

\bibliographystyle{aapmrev4-2}
\bibliography{nemgeo}

\end{document}